%Paper: cond-mat/9211008
%From: powers@mohlsun.physics.upenn.edu (Tom Powers - PN)
%Date: Mon, 16 Nov 92 09:29:10 EST
%Date (revised): Tue, 17 Nov 92 16:56:43 EST

% ``rigid chiral membranes''
% nelson, powers
% print this using plain TeX
\input harvmac

%%%%%%%%%%%%% philmac.tex %%%%%%%%%%%%%
% first a full nine-point font set
%\font\ninerm=cmr9\font\ninei=cmmi9\font\nineit=cmti9\font\ninesy=cmsy9
%\font\ninebf=cmbx9\font\ninesl=cmsl9
%\def\ninepoint{\def\rm{\fam0\ninerm}% switch back to 10-point type
%\textfont0=\ninerm \scriptfont0=\sevenrm \scriptscriptfont0=\fiverm
%\textfont1=\ninei  \scriptfont1=\seveni  \scriptscriptfont1=\fivei
%\textfont2=\ninesy \scriptfont2=\sevensy \scriptscriptfont2=\fivesy
%\textfont\itfam=\nineit \def\it{\fam\itfam\nineit} \def\sl{\fam\slfam\ninesl}
%\textfont\bffam=\ninebf \def\bf{\fam\bffam\ninebf} \rm}

\hyphenation{anom-aly anom-alies coun-ter-term coun-ter-terms
dif-feo-mor-phism dif-fer-en-tial super-dif-fer-en-tial dif-fer-en-tials
super-dif-fer-en-tials reparam-etrize param-etrize reparam-etriza-tion}

%\input usr1:[nelson.texutil]mac.tex

%
% Tagged sections: generates a symbol with current secno and also a
%               table of contents entry
%
% The page number in the toc may be wrong when a section contains no
% text. Try modifying \tnewsec by eliminating \let\the=0 .
%
\newwrite\tocfile\global\newcount\tocno\global\tocno=1
\ifx\bigans\answ \def\tocline#1{\hbox to 320pt{\hbox to 45pt{}#1}}
\else\def\tocline#1{\line{#1}}\fi
\def\toclead{\leaders\hbox to 1em{\hss.\hss}\hfill}
\def\tnewsec#1#2{\xdef #1{\the\secno}\newsec{#2}
\ifnum\tocno=1\immediate\openout\tocfile=toc.tmp\fi\global\advance\tocno
by1
{\let\the=0\edef\next{\write\tocfile{\medskip\tocline{\secsym\ #2\toclead\the
\count0}\smallskip}}\next}% want to expand secsym now, count0 later
}
\def\tnewsubsec#1#2{\xdef #1{\the\secno.\the\subsecno}\subsec{#2}
\ifnum\tocno=1\immediate\openout\tocfile=toc.tmp\fi\global\advance\tocno
by1
{\let\the=0\edef\next{\write\tocfile{\tocline{ \ \secsym\the\subsecno\
#2\toclead\the\count0}}}\next}
}
\def\tappendix#1#2#3{\xdef #1{#2.}\appendix{#2}{#3}
\ifnum\tocno=1\immediate\openout\tocfile=toc.tmp\fi\global\advance\tocno
by1
{\let\the=0\edef\next{\write\tocfile{\tocline{ \ #2.
#3\toclead\the\count0}}}\next}
}
%
% generate rudimentary table of contents
%

%
%
%
% manuscript control
%
% Phys Rev Letters: (those jerks)
% doublespace, put footnotes among references, doublespace refs, set flag

\gdef\prlmode{F}
\long\def\optional#1{}
\def\cmp#1{#1}         %or remove for stuffy journals
%
%   Matters of taste
%
\let\narrowequiv=\equiv
\def\equiv{\;\narrowequiv\;}

     %or else other way round
\fontdimen16\tensy=2.7pt\fontdimen17\tensy=2.7pt %an experiment
 %usual dup not needed with the above
%\mathsurround=1pt %screws up \dsl!

% for the drafts this is useful: (remember not to do \listrefs)
%\def\ref#1#2{\edef\tmp{$[$\string#1$]$} \tmp\edef#1{\tmp}}

% for reduction this is useful:
% \ifx\answ\bigans ...<some equation> \else ... <same with linebreaks> \fi

%
%Greek abbreviations
\def\ga{\gamma}

\def\ep{\epsilon}
\def\al{{\alpha}}

%
%Curly letters
%

\def\CO{{\cal O}}

%
%
%   Miscellaneous
%
% Usage: \boxit{3.5}{some text}
\def\boxit#1#2{
        $$\vcenter{\vbox{\hrule\hbox{\vrule\kern3pt\vbox{\kern3pt
	\hbox to #1truein{\hsize=#1truein\vbox{#2}}\kern3pt}\kern3pt\vrule}
        \hrule}}$$
}

%\def\bivector#1{{\buildrel \leftrightarrow\over #1}}

         % |#1>
         % <#1|
 %matrix element <#1|#2|#3>
%

\def\lfr#1#2{{\textstyle{#1\over#2}}} %       little fraction
 %     tensor up-down

%\def\THETA#1#2#3{\vartheta \hbox{${#1\atopwithdelims[]#2}$}
%               \bigl(#3|\tau\bigr)}
%\def\THETB#1#2#3#4{\vartheta \hbox{${#1\atopwithdelims[]#2}$}
%               \bigl(#3|#4\bigr)}

             % no dot i

% split exact sequence
\def\splitexact#1#2{\mathrel{\mathop{\null{
\lower4pt\hbox{$\rightarrow$}\atop\raise4pt\hbox{$\leftarrow$}}}\limits
^{#1}_{#2}}}

%semi-direct product |><
%
%   Various delbars
%
\def\pa{\partial}

            %delbar dagger
              %delbar dagger delbar
  %delbar n dagger delbar n
            %del-bar-sub n
           %delbar
 %partial derivative
 %partial derivative with room for tilde etc
     %d'alembertian
%
%   Other things with bars
%
          % Beltrami
 % conjugate Beltrami
  % z-bar
  % q-bar
  % tau-bar
\def\ub{{\bar{\vphantom\i u}}}  % u-bar
  % a-bar
      % mu-bar
%
%   Various romans for math
%

  % Im, Re
                 % exponential
\def\dd{\mskip 1.3mu{\rm d}\mskip .7mu} % exterior derivative
                   % dz dzbar
                      % trace
%\font\llcaps=cmcsc10
        % DET bundle
        % DET bundle

                % det
                % sdet

%
%   Whole words
%

\def\IM{isomorphism}

\def\eg{{\it e.g.}}\def\ie{{\it i.e.}}\def\etc{{\it etc.}}

%
%   Fonts
%
%\font\CAPS=cmcsc10 scaled 1200
%\def\cmfontflag{cm}
%\ifx\fontflag\cmfontflag\font\CAPS=cmcsc10 scaled 1200
% \ifx\answ\bigans\font\speci=bbb12 scaled 833\else\font\speci=bbb10\fi
%\else
% \font\CAPS=amcsc10 scaled 1200
%\fi
%\font\blackboard=msym10 \font\blackboards=msym7   % cool
%\font\blackboardss=msym5
%\font\blackboard=cmssbx10 \font\blackboards=cmssbx10 at 7pt  % wimpy
%\font\blackboardss=cmssbx10 at 5pt
%\newfam\black
%\textfont\black=\blackboard
%\scriptfont\black=\blackboards
%\scriptscriptfont\black=\blackboardss
%\def\blackb#1{{\fam\black\relax#1}}
%\def\spec#1{\hbox{\speci #1}}             %historical
%\let\spec=\blackb
%
%\font\gothic=eufm10 \font\gothics=eufm7
%\font\gothicss=eufm5
%\font\gothic=cmssbx10 \font\gothics=cmssbx10 at 7pt  % wimpy substitute
%\font\gothicss=cmssbx10 at 5pt
%\newfam\gothi
%\textfont\gothi=\gothic
%\scriptfont\gothi=\gothics
%\scriptscriptfont\gothi=\gothicss
%\def\goth#1{{\fam\gothi\relax#1}}
%
%\font\curlyfont=eusm10 \font\curlyfonts=eusm7
%\font\curlyfontss=eusm5
%\newfam\curly
%\textfont\curly=\curlyfont
%\scriptfont\curly=\curlyfonts
%\scriptscriptfont\curly=\curlyfontss
%\def\cal{\fam\curly\relax}            %for example

%\font\bbol=cmbx10 scaled \magstep1    %chapter titles
%\font\df=cmssbx10                      %good for definitions

%macros to get automatically-numbered figures
\global\newcount\figno \global\figno=1
\newwrite\ffile
\def\pfig#1#2{Fig.~\the\figno\pnfig#1{#2}}
\def\pnfig#1#2{\xdef#1{Fig. \the\figno}%
\ifnum\figno=1\immediate\openout\ffile=figs.tmp\fi%
\immediate\write\ffile{\noexpand\item{\noexpand#1\ }#2}%
\global\advance\figno by1}

% This one embeds figs in the text (don't mix \tfig with \pfig!)
% Unlike \pfig, this one comes in two parts: \tfig allocates the number and
% \ifig inserts the empty box
% e.g. blah, blah (\tfig\foo) blah, end of paragraph.
%
% \ifig\foo{caption.}{2.5}
%
% New paragraph; a box 2.5truein has been left and captioned.
\def\tfig#1{Fig.~\the\figno\xdef#1{Fig.~\the\figno}\global\advance\figno by1}
\def\ifig#1#2#3{\midinsert\vskip #3truein\narrower\narrower\noindent#1:
#2\endinsert}

% new stuff for bozo

%\def\hal{{1\over2}} %obsolete

%\def\sump{\CO({\textstyle{\sum_{i=1}^q}}Q_i-{\textstyle{\sum_{j=1}^q}}P_i)}

%%%%%%%%%%%%%% goodies for super riemann surfaces %%%%%%%%%%%%%%

 %beltrami diffl

   % big D

                     % bold z
                     % bold u
  % bold z-bar
  % bold u-bar

%
% A-hats
   %curly A-hat
   %curly A-hat, holo
   %curly A-hat, antiholo
%
% Omega-hats

%
% O-hats

%%%%%%%%%%%%%%%%%%%%%%%%%%%%%%%%%%%%%%%%

% Poor man's Blackboard Bold characters often used :
\def\inbar{\,\vrule height1.5ex width.4pt depth0pt}
\def\IB{\relax{\rm I\kern-.18em B}}
\def\IC{\relax\hbox{$\inbar\kern-.3em{\rm C}$}}
\def\ID{\relax{\rm I\kern-.18em D}}
\def\IE{\relax{\rm I\kern-.18em E}}
\def\IF{\relax{\rm I\kern-.18em F}}
\def\IG{\relax\hbox{$\inbar\kern-.3em{\rm G}$}}
\def\IH{\relax{\rm I\kern-.18em H}}
\def\II{\relax{\rm I\kern-.18em I}}
\def\IK{\relax{\rm I\kern-.18em K}}
\def\IL{\relax{\rm I\kern-.18em L}}
\def\IM{\relax{\rm I\kern-.18em M}}
\def\IN{\relax{\rm I\kern-.18em N}}
\def\IO{\relax\hbox{$\inbar\kern-.3em{\rm O}$}}
\def\IP{\relax{\rm I\kern-.18em P}}
\def\IQ{\relax\hbox{$\inbar\kern-.3em{\rm Q}$}}
\def\IR{\relax{\rm I\kern-.18em R}}
%\font\cmss=cmss10 \font\cmsss=cmss10 at 10truept%!!! should be 7pt
\def\IZ{\relax\ifmmode\mathchoice
{\hbox{\cmss Z\kern-.4em Z}}{\hbox{\cmss Z\kern-.4em Z}}
{\lower.9pt\hbox{\cmsss Z\kern-.36em Z}}
{\lower1.2pt\hbox{\cmsss Z\kern-.36em Z}}\else{\cmss Z\kern-.4em Z}\fi}
\def\IGa{\relax\hbox{${\rm I}\kern-.18em\Gamma$}}
\def\IPi{\relax\hbox{${\rm I}\kern-.18em\Pi$}}
\def\ITh{\relax\hbox{$\inbar\kern-.3em\Theta$}}
\def\IOm{\relax\hbox{$\inbar\kern-3.00pt\Omega$}}

%\prlformat\def\cmp#1{}
\def\figcheck{y}
\def\figmode{y}  % suppresses figures
%\long\def\optional#1{\par\hrule {\bf #1}\hrule }

\def\prlcheck{T}
\let\epsilon=\varepsilon
\ifx\answ\bigans \else\noblackbox\fi
%\font\caps=cmcsc10
\Title{\vbox{\hbox{UPR--518T}}}{Rigid Chiral Membranes}
\def\eff{_{\rm eff}}
\def\cost{{c_0^*}}
\def\dxi{\dd^2\xi}
\def\ceff{{c^*\eff}}
\def\ubb{\ub_{12}}
\def\ko{{\kappa_0}}

\centerline{Philip Nelson and Thomas Powers}\smallskip
\centerline{Physics Department, University of Pennsylvania}
\centerline{Philadelphia, PA 19104 USA}
\bigskip\bigskip

Statistical ensembles of flexible two-dimensional
fluid membranes arise naturally
in the description of many physical systems. Typically one encounters such
systems in a regime of low tension but high stiffness against bending, which
is just the opposite of the regime described by the Polyakov string.
We study a
class of couplings between membrane shape and in-plane order
which break 3-space parity invariance. Remarkably there
is only {\it one} such allowed
coupling (up to boundary terms);
this term will be present for any lipid bilayer
composed of tilted chiral molecules. We calculate the renormalization-group
behavior of this relevant coupling in a simplified model
and show how thermal fluctuations effectively
reduce it in the infrared.

%\draft   \nolabels                                 %%% later delete this
\Date{8/92.}\noblackbox                 %%% and reinstate this

\lref\NePelc{D.R. Nelson and R. Pelcovits,
\cmp{``Momentum-shell recursion relations, anisotropic spins, and liquid
crystals in $2+\epsilon$ dimensions,''}
Phys. Rev. {\bf B16} (1977) 2129.}
\lref\Orlando{O. Alvarez, \cmp{``Theory of strings with boundary,''} Nucl.
Phys. {\bf B216} (1983) 125.}
\lref\Brow{E. Browicz, Zbl. Med. Wiss. {\bf28} (1890) 625.}
\lref\HPP{G. Hinshaw, R. Petschek, R. Pelcovits, \cmp{``Modulated
phases in thin ferroelectric liquid-crystal films,''}
Phys. Rev. Lett. {\bf60} (1988) 1864;
G. Hinshaw and  R. Petschek, \cmp{``Transitions and modulated phases in chiral
tilted smectic liquid crystals,''}
Phys. Rev. {\bf A39} (1989) 5914.} %11/16/88
\lref\LaSe{S. Langer and J. Sethna, \cmp{``Textures in a chiral smectic
liquid-crystal film,''}
Phys. Rev. {\bf A34} (1986) 5035.} %3/10/86
\lref\Meun{J. Meunier, \cmp{``Liquid interfaces,''}
J. Phys. (Paris) {\bf 48} (1987) 1819.} %3/11/87.
\lref\SSRNO{R. Strey {\it et al.}, \cmp{``Dilute Lamellar and L$_3$
phases in the binary water--C$_{12}$E$_5$ system,''}
J. Chem. Soc. Faraday Trans. {\bf 86}(12) (1990) 2253.} %12/12/89.
\lref\HelPro{W. Helfrich and J. Prost, \cmp{``Intrinsic bending force
in anisotropic membranes made of chiral molecules,''}
Phys. Rev. {\bf A38} (1988) 3065; %.} %3/23/88
Ou-Yang Zhong-Can and Liu Jixing, \cmp{``Theory of helical structures
of tilted chiral lipid bilayers,''}
Phys. Rev. {\bf A43} (1991) 6826.} %11/26/90
\def\OYL{}
\lref\DiPiMe{S. Dierker, R. Pindak, and R. Meyer, Phys. Rev. Lett.
{\bf 56} (1986) 1819; W. Heckl {\it et al.,} Eur. Biophys. J. {\bf14} (1986)
11.}
\let\pinwheel=\DiPiMe
\lref\Helfb{W. Helfrich, \cmp{``Helical bilayer structures due to spontaneous
torsion of the edges,''}
J. Chem. Phys. {\bf 85}(2) (1986) 1085.} %3/17/86
\lref\DaLe{F. David and S. Leibler, \cmp{``Vanishing tension of fluctuating
membranes,''} J. Phys. II France {\bf 1} (1991) 959.} %1/3/91
\lref\SMMS{See for example {\sl Statistical mechanics of membranes and
surfaces,} D. Nelson {\it et al.}, eds (World Scientific, 1989).}
\lref\DSMMS{%
A. Polyakov, {\sl Gauge fields and strings}, (Harwood, 1987);
F. David, in \SMMS.}
\def\Polbook{}
\lref\CaHe{P. Canham , J. Theor. Biol. {\bf26} (1970) 61; W.
Helfrich, Naturforsch. {\bf28C} (1973) 693.}
\lref\DaLa{I. Dahl and S. Lagerwall,
Ferroelectrics {\bf 58} (1984) 215;
L. Peliti and J. Prost, \cmp{``Fluctuations in membranes with reduced
symmetry,''} J. Phys. France {\bf 50} (1989) 1557; %.} %12/5/88
T.C. Lubensky and F. MacKintosh, \cmp{``Orientational order, topology,
and vesicle shapes,''} Phys. Rev. Lett. {\bf67} (1991) 1169.}

\lref\Ratpriv{B. Ratna, private communication.}
\lref\JPAS{J. Polchinski and A. Strominger, \cmp{``Effective string theory,''}
Phys. Rev. Lett. {\bf67} (1991) 1681.}
\lref\promise{P. Nelson and T. Powers, in preparation.}

%%%%%%%%%%%%%%%%%%%%%%%%%%%%%%%%%%%%%%%%%
Statistical ensembles of random geometrical shapes pervade theoretical physics.
Initially one-dimensional curves in space were most thoroughly studied
due to
their ease of description and the many applications of such ensembles to the
conformation and dynamics of polymers, but today ensembles of two-dimensional
membranes are at least as important. Physical realizations of such surfaces
include lipid bilayers and surfactant films, which spontaneously self-assemble
{}from amphiphilic molecules in solution or at fluid interfaces (for reviews
see \SMMS). More speculative applications as diverse as the 3d Ising model and
other 3d phase transitions, cosmic strings, flux tubes in QCD, and models of
elementary particles all rest upon the key property that the important physical
degrees of freedom in these problems are {\it shapes} with no preferred
choice of coordinates.
%ok
{\foot{Sheets of molecules frozen into fixed lattices
form ``tethered'' membranes
which do have a preferred coordinate
system, much like elastic solids. Here we study only the opposite case
(``fluid'' membranes).
}}\ The
condition that coordinate choice be immaterial greatly constrains the possible
forms of the statistical weights in these systems, leading to very few
independent couplings and hence physically simple models.

In this letter we will study a model appropriate for the description of
tilted
lipid bilayers (\eg\ the lamellar $L_{\beta^*}$ phases of lyotropics or
$S_{C^*}$ phases of smectics), though we think the analysis is potentially
interesting in other contexts as well. Bilayers are typically {\it rigid},
that is their resistance to bending is characterized by a dimensionless
quantity $\bar\kappa_0\equiv\kappa_0/k_BT$ (see below) which is greater than
one.
Accordingly, we will carry out a perturbative expansion about $\bar\kappa_0\to
\infty$, the high-stiffness, low-temperature limit. Bilayers with free boundary
conditions also typically adjust themselves to zero effective surface tension
\DaLe, and we will also work in this limit.%
\nref\NePe{D.R. Nelson and L. Peliti, \cmp{``Fluctuations in membranes with
crystalline and hexatic order,''} J. Physique (Paris) {\bf 48} (1987)
1085.}%12/22/86
\nref\DaGuPe{F. David, E. Guitter, and L. Peliti, \cmp{``Critical
properties of fluid membranes with hexatic order,''} J. Phys. (Paris)
{\bf48} (1987) 2059.}%
\foot{This is opposite to the
usual Polyakov string, in which tension dominates and $\bar\kappa_0$ is
effectively zero. Refs. \NePe\DaGuPe\ have shown in a related model (hexatic
membrane) that with in-plane order
the stiffness can stabilize at a large value without running
to zero at long scales. In any case the high-stiffness limit is appropriate
for a system viewed on scales shorter than its persistence length, if any.
In this paper, we will not study the running of the
various stiffness couplings at all, concentrating instead on the behavior of
chirality.}

We can summarize our logic as follows (further details will appear elsewhere).
At length scales far longer than the size of the constituent molecules a
continuum description becomes appropriate. In our nearly flat
regime the important
degrees of freedom are the elastic (or ``Goldstone'') modes corresponding to
transverse undulations as well as
director fluctuations if in-plane order develops.
We will find only a few allowed couplings of these modes, as we expect in any
elastic system at very long wavelengths: the system forgets most of the
details about its constituents. Somewhat surprisingly, however, the membranes
with in-plane order {\it can} remember at long scales whether or not their
constituent molecules are {\it chiral}, even though this chirality is often
a rather subtle property of the amphiphiles. Gross chiral behavior has long
been seen in monolayers, where one gets pinwheel domains \pinwheel,%
%!!!
{\foot{We must be careful not to confuse the explicit
breaking of parity symmetry due to chiral molecules with the {\it
spontaneous} breaking of parity symmetry
(R. Bruinsma and J.
Selinger, to appear)
seen in pinwheel domains made from {\it achiral} molecules
(X. Qiu {\it et al.}, Phys. Rev. Lett. {\bf67} (1991) 703%.}
).}}
 as well
as in flexible membranes which can form helical ribbons~%
\ref\ribbons{N. Nakashima {\it et al.}, \cmp{``Helical superstructures
are formed from chiral ammonium bilayers,''} Chem. Lett. {\bf1984}
(1984) 1709; K. Yamada {\it et al.}, \cmp{``Formation of helical super
structure from single-walled bilayers by amphiphiles with
oglio-L-glutamic acid head group,''}
Chem. Lett. {\bf1984} (1984) 1713;
J. Georger {\it et al.}, \cmp{``Helical and tubular microstructures
formed by polymerizable phosphatidylcholines,''}
J. Am. Ch. Soc.  {\bf109} (1987)
6169.}
%!!!!
whose sense depends on the constituent molecules' handedness~%
\ref\Nakashima{N. Nakashima {\it et al.}, in \ribbons;
A. Singh {\it et al.}, \cmp{``Lateral phase separation based on
chirality in a polymerizable lipid and its influence on formation of
tubular microstructures,''} Chem. Phys. Lipids {\bf47} (1988) 135.}%
. We will see that this memory follows
{}from the existence of an allowed parity-violating term in the free energy,
which couples in-plane order to shape.%
\ref\Rhodesetal{In a closely related system
(microtubules) there is experimental evidence that in-plane tilt order is
indeed present.
For example  see
R. Treanor and M. Pace, \cmp{``Microstructure, order, and fluidity of
a polymerizable lipid by ESR and NMR,''} Biochim. Biophys. Acta {\bf1046}
(1990) 1;
B. Thomas {\it et al.}, \cmp{``X-ray diffraction studies of tubules
formed from a diacetylenic phosphocholine lipid,''} in {\sl Complex
fluids,} Mat. Res. Soc. Symp.  Proc. {\bf248} (1992) 83.}%
{} When in-plane order is thermally destroyed the theory admits
no such terms;
the system
cannot express the chirality of its constituents even if present. Indeed,
experimentally a loss of gross chiral structure does seem to accompany
the chain-melting transition, and chiral structures do not form
at all above
this temperature~\ref\YaSh{P. Yager and P. Schoen,
\cmp{``Formation of tubules by a polymerizable surfactant,''} Mol. Cryst.
Liq. Cryst. {\bf106} (1984) 371.}%
\Nakashima.
%!!!!!!!

Thermal fluctuations are often important in membranes at room temperature
where $\bar\kappa_0\,\roughly<\, 40$ is not too close to infinity. It is well
known that without in-plane order such fluctuations induce a logarithmically
scale-dependent softening of the effective stiffness $\bar\kappa$ in the
infrared
\ref\DeTa{P.G. de\thinspace Gennes and C.  Taupin, J. Phys. Chem.
{\bf86} (1982) 2294;
W. Helfrich, J. Phys. (Paris) {\bf46} (1985) 1263; {\it
ibid.} {\bf47} (1986) 321;
L. Peliti and S. Leibler, \cmp{``Effects of thermal fluctuations on
systems with small surface tension,''}
Phys. Rev. Lett. {\bf54} (1985) 1690; %2/4/85
D. F\"orster, \cmp{``On the scale dependence due, to thermal
fluctuations, of the elastic properties of membranes,''}
Phys. Lett {\bf114A} (1986) 115;
A. Polyakov, \cmp{``Fine structure of strings,''}
Nucl. Phys. {\bf B268} (1986) 406; %10/21/85
%\lref\Kla{
H. Kleinert, \cmp{``Thermal softening of curvature elasticity in
membranes,''} Phys. Lett. {\bf 114A} (1986) 263.}%11/14/85
%\Helfa\PeLe\Fors\Pold
. What we will show is that the
unique bulk chiral coupling, if present,
similarly suffers a logarithmic renormalization%
{} The RG behavior of this chiral term seems not to have been studied
before.
While some of our analysis will reproduce others' results, we
hope that our unified treatment of allowed free energy terms will clarify
some of the important symmetries; we have also tried to clear up several
subtleties in the fluctuation problem~\promise.
Finally, the renormalization of chirality will affect the average shapes
taken on by membranes, since the chiral coupling helps determine those
shapes. We will derive an anomalous scaling relation for the
radius of helical ribbons as chirality varies  which departs from the
mean-field formula~\HelPro\OYL\ and may be experimentally
testable.\foot{We thank J.~Toner for suggesting this consequence.}

We begin by considering in greater detail the elastic modes of our system.
A membrane made of molecules which slip around each other in some average
2d locus with no in-plane order at all can be described just by specifying
a mathematical 2-surface $\vec x (\xi)$ in 3-space. We can choose coordinates
$\xi^i$, $i=1,2$ for our surface, but we must remember that this choice
is arbitrary.
The choice of a flat reference surface breaks one transverse translation
symmetry, giving one ``undulation'' mode. We can visualize this mode as
height fluctuations from a horizontal plane. At high temperatures this is
the only soft mode we expect. At lower temperatures in-plane order can
develop, leading to additional soft modes. While experiments do not seem
to see full crystalline in-plane order in (hydrated, unpolymerized) membranes%
{}~\Rhodesetal, and such order is
disfavored on theoretical grounds~\NePe,
still
orientational quasi-long-range order can survive,
giving rise to an {\it angular} elastic mode. We will consider only
temperatures well within the ordered phase. Hence we can
visualize the corresponding order parameter
as a unit vector field $\vec m(\xi)$ {\it tangent}
to our 2-surface.

For hexatic in-plane order $\vec m$ is defined only up to rotations by
$2\pi/6$. When the constituent molecules are tilted from the surface normal
they break rotation invariance completely, so $\vec m$ has no periodic
identifications, but a new subtlety arises instead. To define $\vec m(\xi)$
given a configuration of molecules, we must project the molecule axis down
to the midplane of the membrane. Since the two sides of the bilayers are
equivalent, the overall sign of $\vec m(\xi)$ is not fixed until we {\it
choose} one side, or equivalently, one of the two normal vectors $\hat n(\xi)$;
with this choice we may use the convention where the projection of the molecule
on the outward-facing layer defines $\vec m$%
\ifx\figmode\figcheck\else{} (\tfig\fone)\fi.
Since there is no preferred choice
of normal, each term of our free energy must be unchanged under the
substitution $\vec m\mapsto-\vec m$, $\hat n\mapsto-
\hat n$.\foot{For monolayers or closed vesicles there {\it is} a preferred
choice of normal and we should not impose this symmetry.}

\ifx\figmode\figcheck\else
\ifig\fone{Ambiguity in defining the in-plane order parameter: $(\hat n
,\vec m\,)$ describe the same configuration as $(\hat n',\vec m')$.}{1.7}
\fi

Our surface gets a few standard tensor fields. It inherits a metric $g_{ij}=
\partial_i\vec x\cdot\partial_j\vec x$,
a corresponding %Levi-Civita
covariant derivative $\nabla$,
and a volume form $\dd^2\xi\sqrt{g}$. Once we choose a normal $\hat n$, we
also get a second fundamental form $K_{ij}\equiv\hat n\cdot\nabla_i\partial_j
\vec x$\optional{, an induced orientation,}{} and an alternating tensor
$\ep_{ij}\equiv
\ep_{ija}\hat n^a =
\sqrt{g}\,{\ninepoint\pmatrix{0&1\cr-1&0\cr}}_{ij}$. Here
$\ep_{ija}\equiv\partial_i x^b\partial_j x^c\ep_{bca}$ where $\ep_{bca}$ is
the usual alternating symbol, $\ep_{123}=+1$. Note that $\ep_{ij}$ and
$K_{ij}$ change sign under the change of normal
$\hat n\mapsto-\hat n$, $\vec m\mapsto-\vec m$ discussed above.

Let us enumerate all allowed free energy terms. We assume all
nonlocal interactions are absent or screened to a length scale shorter than
the scale of interest. Then we simply seek all local terms in $\vec x(\xi)$
and $\vec m(\xi)$ with the above symmetries, relevant or marginal in the
low-temperature or high-stiffness expansion. We will not seek to derive
such terms from some 3-dimensional liquid crystal free energy, but instead
simply construct them directly from the ingredients listed above.
The relevance of an operator about the weakly fluctuating $\bar\kappa_0\to
\infty$ fixed point depends on its naive engineering dimension. The only
short-distance cutoff in the problem is the characteristic size $\Lambda^{-1}$
of the constituents. $\Lambda^{-1}$ is a 3-space length, not a parameter-space
length as in ordinary 2d quantum field theory, so when counting the dimensions
of operators we should consider only their behavior under rescaling $\vec x$
(not $\xi^a$).{\foot{One can reproduce this conclusion by carrying
out the traditional analysis of divergences of Feynman diagrams. This power
counting and the nature of the corresponding continuum limit differ from the
procedure used in the Polyakov string where \eg\ $x^a(\xi)$ are considered
as three scalar fields and hence dimensionless.}}\ $\hat n$ and $\vec m$ are
unit vectors and hence dimensionless.
\optional{Thus, for example, $\int\sqrt{g}\,\dd^2
\xi$ has dimensions $L^2$ and so is relevant, while $\int\dd^2\xi\sqrt{g}
\,(g^{ij} K_{ij})^2$ is dimensionless and hence marginal.}

We can now list all the independent relevant and marginal terms allowed in
the free energy:
\eqnn\eHone\eqnn\eHtwo\eqnn\eHstar
$$\eqalignno{
H_1=&{\kappa_0\over2}\,\int\dd^2\xi\sqrt{g}\,
\left[(K^i_{\ i})^2
+\gamma_1(\nabla_i m^j)(\nabla^i m_j)
+\gamma_2(\vec m\cdot\vec\nabla m^i)(\vec m\cdot\vec\nabla m_i)
\right]\quad,&\eHone\cr
%\eqn\eHtwo{\eqalign{
H_2=&{\kappa_0\over2}\,\int\dd^2\xi\sqrt{g}\,
\left[\alpha_1\vec m\cdot K\cdot K\cdot\vec m
+\alpha_2(\vec m\cdot K\cdot\vec m)(K_{\ i}^i) \right.\cr
&\qquad\quad+\alpha_3(\vec m\cdot K\cdot\vec m)^2
+\beta_1(\vec m\cdot K\cdot\vec m)(\vec\nabla\cdot\vec m)
%&\qquad\qquad
+\beta_2(K_{\ i}^i)\vec\nabla\cdot\vec m\cr
&\qquad\quad\left.+\beta_3 K_{\ j}^i\nabla_im^j
+\beta_4 m^\ell K_{\ell i} m^j \nabla_j m^i\right]
\quad,&\eHtwo\cr
%\eqn\eHstar{
H_*=&{c_0^*\over 2}\int\dd^2\xi \,\sqrt g
m^i\ep_i^{\ j}K_{j\ell}m^\ell\quad.&\eHstar\cr}$$
In these formulas we raise and lower $i,j$ indices using $g_{ij}$.
Most of these terms have already been discussed by Helfrich and Prost~\HelPro;
see also \DaLa%\PePr\ref\TCLFM{}%
. $H_1$ is the usual Canham-Helfrich elastic energy \CaHe\ with
zero tension\DaLe,
plus the covariant form of an X--Y model energy. $H_2$ contains
various anisotropies in the bending energy due to the tilt of the constituents.
$\gamma_i,\alpha_i,\beta_i$ are dimensionless numbers we take to be $\CO(1)$,
while $\kappa_0$ is an energy scale we take much larger than the temperature
$T$ (we set Boltzmann's constant = 1). Every term of $H_{1,2}$ (every
nonchiral term) is marginal. $H_*$ is the only allowed bulk chiral term;
it is relevant. As mentioned earlier, there are {\it no} available
relevant or marginal bulk
chiral terms involving only shape (no tilt).

In the enumeration \eHone--\eHstar\ we have dropped all total derivatives,
including for instance the Gaussian curvature and
the covariant form of
the $\vec\nabla\times\vec m$ term \LaSe.
Such terms are important
for systems with boundaries (\eg\ ribbons) or defects (\eg\ rippled phases),
but they will not affect our calculation of the renormalization of $c_0^*$.
(Near defects one should also allow $\vec m$ to vary in magnitude \HPP.)
What is remarkable about \eHone--\eHstar\ is that while there are a number of
allowed couplings, still the list is rather short, especially compared to
a $2d$ scalar at its trivial fixed point,
for which infinitely many marginal terms exist.

We will consider a limit in which $\bar\kappa_0\equiv\kappa_0/T$ is large
but $c^*_0/T$ is very small compared to the cutoff. This is reasonable
since typically chirality is a very minor feature of the long chain
amphiphiles.
Thus, we will carry all our calculations out to lowest nontrivial order in
the loop-counting parameter $(\bar\kappa_0)^{-1}$ and the chiral coupling
$c^*_0$.

We will calculate the effects of thermal fluctuations in saddle-point
(one-loop) approximation. Before we can do the required integrals, however, we
must fix certain coordinate-choice redundancies in our description of
a surface.
Our problem is that we have five variables $\vec x(\xi)$, $\vec m(\xi)$ in
\eHone--\eHstar, while as we have seen only two are truly independent. We
must first write $\vec x$ in terms of one independent field $u(\xi)$. For
our purposes, the easiest choice is ``Monge gauge:''
$\vec x(\xi)=(\xi^1,\xi^2,u(\xi))$,
since nearly-flat surfaces can be expanded easily in powers of $u$. Next we
may write $\vec m$ as
$\vec m(\xi)=\vec e_1(\xi)\cos\theta(\xi)+
\vec e_2(\xi)\sin\theta(\xi)$,
where $\{\vec e_1,\vec e_2\}$ are a field of orthonormal tangent vectors
to the surface. $\vec e_\al$ depend on $\vec x(\xi)$, but $\vec x$ does not
fully determine them: we must fix an $O(2)$ gauge freedom. Monge gauge has
the pleasant feature that we may choose $\vec e_\alpha=e_\alpha^{\ i}\partial_i
\vec x$, where
$e_\alpha^{\ i}=\delta_\alpha^i-\half\partial_iu\partial_\alpha u+
\CO(u^3)$.
This expression is not covariant because %\eMonge\
Monge gauge is not. {From now on
we will raise and lower indices using the {\it flat} metric $\delta_{ij}$;
all $g_{ij}$ factors will be shown explicitly. Similarly, we convert index
type using $\delta_\alpha^i$; all $e_\alpha^{\ i}$ factors will be shown
explicitly.} The fact that we can choose a frame with no
$\CO(u)$ terms
will make Monge gauge very convenient.

We now have all the necessary ingredients. When viewing the system on a scale
$L\gg\Lambda^{-1}$ we may forget about irrelevant couplings; moreover
all the effects of fluctuations on scales between $\Lambda^{-1}$ and
$(b\Lambda)^{-1}\,\roughly>\,\Lambda^{-1}$ may be summarized by readjusting
the values of our couplings, since we took care to include all allowed
terms and our cutoff respects the symmetries.
Our strategy is to expand $u=\bar u+h$, $\theta=\bar\theta+\zeta$, where
$\bar u,\bar\theta$ have only long-wavelength components, while $h,\zeta$
have only wavenumbers greater than $b\Lambda$.%
{\nref\NePelc{D.R. Nelson and R. Pelcovits,
\cmp{``Momentum-shell recursion relations, anisotropic spins, and liquid
crystals in $2+\epsilon$ dimensions,''}
Phys. Rev. {\bf B16} (1977) 2129.}%
\foot{Since $(b\Lambda)\inv$ is still a very short scale we
are allowed to use spin-wave approximation for $\theta$, even though on
long scales an X--Y model does not have true long-range order. We leave
to the renormalization group the task of summing the infinite string of
logarithms giving the true long-range behavior of $\theta$ ({\it cf.}\/
\NePelc).}}
{} We expand the free energy $H$ in $h,\zeta$ about $\bar u,
\bar\theta$, find the quadratic terms
in $h,\zeta$, and integrate
the fast modes of $h,\zeta$ in Gaussian approximation to get the one-loop
effective ``action'' $H_{\rm eff}[\bar u,\bar\theta]$ (see \eg\ \Polbook
\DSMMS).%
{\foot{In general this $H_{\rm eff}$ does not have the same
functional form as
$H$; in general we need to rescale the fields %$\ub,\ \bar\theta$
to recover the
original relations between terms of different degrees of homogeneity
(as for example in the $O(n)$ nonlinear sigma model~\NePelc).
In our problem
there is no need for an additional field renormalization of $u$.
Intuitively, this is clear
{}from
the Monge gauge condition: a rescaling of $x^3$ would by rotation invariance
entail
rescaling $x^{1,2}$ which would spoil the gauge condition. Detailed
calculation confirms that the nonlinear structure of \eHone--\eHstar\ is
retained if we do not rescale $u$ \promise.}}
Finally we
read off the renormalized chiral coupling $c^*_{\rm eff}$.

To keep our formulas manageable we will truncate our model \eHone--\eHstar,
retaining only the first two terms of $H_1$ and of course $H_*$
(for more details see \promise). We
expect that the isotropic terms retained will give a good qualitative
guide to the effects of fluctuations on chirality. The terms we have
omitted will not be generated to the order we are working, since they
lack an extra $O(2)$ symmetry of the first two terms; nor can the chiral
term induce them to first order in $c_0^*$.

We want to pick off from $H_{\rm eff}$ the renormalized coefficient of
$\half\int\sqrt{\bar g}\, \bar m^i\ep_i^{\ j}\bar K_{j\ell}\bar m^\ell$.
Since this is the only parity violating term,
we pick terms in the expansion of the log which have an odd power of $c_0^*$.
Since we work to lowest order in $c_0^*$, this means we keep exactly one power,
\ie\ we simply renormalize the operator $H_*$.
Using Monge gauge % and \eMonge--\eframe\
we find that
\eqn\none{H_*={\cost\over2} \int\dxi\Bigl[u_{12}+\theta(u_{22}-
u_{11}) + \theta u_\ga(u_1u_{1\ga} - u_2u_{2\ga})
-\half u_\ga(u_1u_{2\ga} + u_2u_{1\ga}) - 2\theta^2u_{12}\Bigr] + \cdots,
}
where $u_\ga\equiv\pa_\ga u\equiv\pa u/\pa\xi^\ga$ \etc\ and
the ellipsis denotes terms with at least five fields or at least three
$\theta$'s. While $H\eff$ is a complicated power series in $\ub,\
\bar\theta$, we can unambiguously determine $\ceff$ by expanding $H\eff$
to first order in $\ub$ and zeroth order in $\bar\theta$, since the first
term breaks parity and does not appear in any of the total derivative terms
dropped in \eHstar. While $u_{12}$ is a total derivative, so that this
term seems to vanish, we can still compute it by giving the coupling
$\cost$ a small fictitious spatial dependence in intermediate stages of
the calculation.

\ifx\figmode\figcheck
We now quote the results of the calculation. Letting
$D\equiv\lfr T{4\pi\ko}\log b\inv$,
$H_{*,\rm eff}$
looks like $H_*$ with $\cost$ replaced by $\cost(1-\lfr{4D}{\ga_1})$.
\else
We now quote the results of the Feynman diagrams in \tfig\ftwo. Letting
$D\equiv\lfr T{4\pi\ko}\log b\inv$, graphs (a,b) contribute
$-\cost D(\half+\lfr 2{\ga_1})\int\dxi\,\ubb$, while (c) contributes
$\lfr12\cost D\int\dxi\,\ubb$. All told then we find that $H_{*,\rm eff}$
looks like $H_*$ with $\cost$ replaced by $\cost(1-\lfr{4D}{\ga_1})$.
\fi
Hence we find that the effect of fluctuations may be summarized by
omitting them but replacing $\cost$ by $\ceff$, where
\eqn\eRG{{\dd \ceff(b\inv)\over\dd\log b\inv}={-\ceff T\over\pi\ko\ga_1}
\quad.}
The constants $\ko, \ga_1$ were defined in \eHone. Strictly speaking
they too should be allowed to run, but as mentioned they will arrive at
\def\keff{\kappa_{\rm eff}}
the fixed line \NePe\ $(\kappa\ga_1)_{\rm eff}=4\keff$.
Solving \eRG\ we find
\eqn\eanom{\ceff(b\inv)=(b\inv)^{-T/4\pi\keff}\cost<\cost\quad,}
the promised thermal softening.

Eqn.~\eanom\ should not be construed as a temperature-dependence of
$\ceff$, since in general the bare couplings will have some unknown
$T$-dependence. However we can draw an interesting qualitative
conclusion. The diameter $R$ of helical ribbons (and possibly tubules as
well) is controlled by a competition between chirality and stiffness.
Oversimplifying somewhat by omitting the $\vec\nabla\times\vec m$ boundary
term (its effect is similar to that of $H_*$ \HelPro), one finds in
mean-field theory that~\HelPro\OYL\ $R\propto\ko/\cost$. Since $\cost$
can be much smaller than the cutoff $\Lambda$, $R$ can be very large;
indeed experimentally $R$ can be $\sim .5\mu$m~\ribbons. Hence thermal
fluctuations can significantly modify the mean-field result; we can
approximately account for their effects~\promise\ by writing $R
\propto\keff/\ceff(\Lambda R)$. Thus \eanom\ says that varying $\cost$
for fixed $\ko,\ \ga_1,\ T$ we have that the cylinder radius scales not
as $R\propto(\cost)\inv$ but as
\eqn\efinal{R\propto(\cost)^{-(1+T/4\pi\keff)}\quad.}
The nice feature of \efinal\ is that it may be possible experimentally
to control $\cost$, without changing significantly the other
bare parameters, simply by diluting the chiral amphiphiles with similar
but achiral analogs~\Ratpriv. Thus \efinal\ is potentially a rather
clean test of renormalization effects in rigid chiral membranes. Whether
the range of dilutions admitting helices or tubules will be great
enough, and $T/4\pi\keff$ can be made large enough, to test the scaling
law \efinal\ remains to be seen. Even if not, \eanom\ may still be
applicable to rippled phases~\HelPro.

\ifx\figmode\figcheck\else
\ifig\ftwo{Diagrams entering the renormalization of $\cost$. The star denotes
vertices coming from $H_*$.}{1.2}
\fi

We have seen how the constituents of a membrane can express their chiral nature
at long scales through the development of in-plane tilt order, and how
thermal fluctuations can reduce the effective value of the bulk chiral
coupling constant and in turn affect the shapes of self-assembled
structures.
Near the trivial,
low temperature, fixed point the chiral term is relevant, but we have
seen how thermal fluctuations reduce its effective dimension. This raises
the possibility of  a {\it critical} chiral membrane when this term becomes
marginal. Unfortunately this will not happen at weak coupling, so we can say
little about this intriguing possibility.

%%%%%%%%%%%%%%%%%%%%%%%%%%%%%%%%%%%%%%%%%
\ifx\prlmode\prlcheck\bigskip\else
\vskip1.1truein \leftline{\bf Acknowledgements}\fi
{\frenchspacing\noindent
We would like to thank R. Bruinsma,
F. David, M. Goulian, E. Guitter,
F. MacKintosh,
S. Milner,
D.R. Nelson,
J. Polchinski,
J. Preskill,
B. Ratna,
C. Safinya,
J. Schnur,
R. Shashidar,
E. Wong,
and especially S. Amador, T. Lubensky, and J. Toner for innumerable
discussions.} This work was supported in part by NSF grant PHY88-57200, the
Petroleum Research Fund, and the A.~P.~Sloan Foundation. P.N.~thanks
the Aspen Center for Physics for hospitality while this work was being
completed.
\goodbreak
\listrefs
\bye